\documentclass[pdflatex,sn-basic]{sn-jnl}% Basic Springer Nature Reference Style/Chemistry Reference Style

\usepackage{epsf}
\usepackage{graphicx}% Include figure files
\usepackage{dcolumn}% Align table columns on decimal point
\usepackage{bm}% bold math
\usepackage{subfigure}
\usepackage{color}
\usepackage{amsthm}
\usepackage{amssymb}
\usepackage{amsmath}
\usepackage{pb-diagram}
\usepackage{epstopdf}
\usepackage[thinc]{esdiff}

\jyear{2022}%

\theoremstyle{thmstyleone}%
\theoremstyle{thmstyletwo}%

\theoremstyle{thmstylethree}%

\begin{document}

\title[Speciation in a MacArthur model...]{Speciation in a MacArthur model predicts growth, stability and adaptation in ecosystems dynamics.}

%%=============================================================%%
%% Prefix	-> \pfx{Dr}
%% GivenName	-> \fnm{Joergen W.}
%% Particle	-> \spfx{van der} -> surname prefix
%% FamilyName	-> \sur{Ploeg}
%% Suffix	-> \sfx{IV}
%% NatureName	-> \tanm{Poet Laureate} -> Title after name
%% Degrees	-> \dgr{MSc, PhD}
%% \author*[1,2]{\pfx{Dr} \fnm{Joergen W.} \spfx{van der} \sur{Ploeg} \sfx{IV} \tanm{Poet Laureate} 
%%                 \dgr{MSc, PhD}}\email{iauthor@gmail.com}
%%=============================================================%%

\author[1]{\fnm{Elena} \sur{Bellavere}}%\email{iauthor@gmail.com}

\author[2]{\fnm{Christian H.S.} \sur{Hamster}}\email{christian.hamster@wur.nl}

\author[1]{\fnm{Joshua A.} \sur{Dijksman}}\email{joshua.dijksman@wur.nl}

\affil*[1]{\orgdiv{Physical Chemistry and Soft Matter}, \orgname{ Wageningen University}, \orgaddress{\street{Stippeneng 4}, \city{Wageningen}, \postcode{6708 WE}, \country{Netherlands}}}

\affil[2]{\orgdiv{Biometris}, \orgname{Wageningen University}, \orgaddress{\street{Droevendaalse steeg 1}, \city{Wageningen}, \postcode{6708 PB}, \country{Netherlands}}}

%%==================================%%
%% sample for unstructured abstract %%
%%==================================%%

\abstract{Ecosystems dynamics is often considered as driven by a coupling of species' resource consumption and its population size dynamics. Such resource-population dynamics is captured by MacArthur-type models. One biologically relevant feature that would also need to be captured by such models is the introduction of new and different species. Speciation introduces a stochastic component in the otherwise deterministic MacArthur theory. We describe here how speciation can be implemented to yield a model that is consistent with current theory on equilibrium resource-consumer models, but also displays readily observable rank diversity metric changes. The model also reproduces a priority effect. Adding speciation to a MacArthur-style model so provides an attractively simple extension to explore the rich dynamics in evolving ecosystems.}


%%\pacs[JEL Classification]{D8, H51}

%%\pacs[MSC Classification]{35A01, 65L10, 65L12, 65L20, 65L70}

\maketitle

\section{Introduction}
An ecosystem is a set of species, each of finite population size that interact by competing for finite resources that fuel their growth. A single ecosystem can involve dynamics that occurs over a wide range of length and timescales~\cite{azaele2016}; Darwin already eloquently referred to this in his ``tangled bank'' remark. The seemingly universal nature of a species' emergence, adaptation and extinction in such ecosystems, has inspired many to describe the phenomenology of ecosystem dynamics with simple modeling with only a few ingredients that are independent of the specific physical mechanisms at play~\cite{nowak2006, azaele2016}. What is then the simplest quantitative description that displays all the salient dynamical features of evolution? This question has a long list of partial answers~\cite{nowak2006,tikhonov2016community,posfai2017metabolic}, although much work in the field concerns equilibria~\cite{macarthur1963,may1972,chesson1990,hubbell2001unified}. Here we show that the already successful variants of MacArthur models can be amended with a simple stochastic mechanism that introduces new species, which allows us to show many of the biologically relevant \emph{dynamical} features of evolution even when the starting point for the dynamics is a single primordial ancestor. In particular, our MacArthur model variant can describe the growth dynamics of an ecosystem in terms of species richness, its evolution towards a dynamic equilibrium size, and even its adaptation to resource influx changes. The predictions of the model are consistent with other existing modeling on for example equilibrium dynamics, and reasonably in line with common observations on for example resource shock experiments.\\

\section{The speciating MacArthur model}
Evolutionary dynamics modeling including MacArthur type models usually start with describing population dynamics as $\dot{\textbf{n}}(t) = \textbf{n}(t)f(\textbf{\textbf{n}})$ with $\textbf{n}$ the set of species population sizes and the dot denotes a time derivative. The crux is that $f(\textbf{n})$ is not constant but a growth rate determining \emph{function} that depends on ecosystem features such as population sizes and coupling constants which specify inter-species competition and preying efficiency~\cite{wangersky1978, cressman2014}. This general approach is tremendously successful even in capturing quantitative experimental observations of low dimension systems~\cite{korolev2011b}. However, describing the dynamics of larger ecosystems with many evolving species is challenging, as high dimensional systems quickly lose their numerical and analytical tractability, even without incorporating the additional complexity of the evolution of each species.

We address this complexity by adding evolutionary dynamics to multi-species ecosystem models by quantifying the growth rate within the context of the ecosystem properties. It has become customary in recent years to \emph{define} every species $j$ by a \emph{strategy vector} $\textbf{s}_j$~\cite{posfai2017metabolic, tikhonov2017collective, pacciani2020dynamic, caetano2021evolution} that couples the growth dynamics $\dot{\textbf{n}}(t) = \textbf{n}(t)f(\textbf{\textbf{n}})$ to a dynamic resource vector $\textbf{r}(t)$ which represents the amount of available resources at time $t$. Here, each component $i$ of $\textbf{s}_j$ describes which fraction of each resource $r_i$ is used by every species $j$ at every time step. The time-dependent growth factor is then naturally captured by the alignment $\textbf{s}_j\cdot\textbf{r}$. We can write for each element $\dot{\textbf{n}}_j = \textbf{n}_jf_j(\textbf{n},\textbf{s}_j\cdot\textbf{r})$ while introducing resource time-dependence via a function $\dot{\textbf{r}} = g(\textbf{n},\textbf{s})$, where $\textbf{s}$ is the matrix with strategies $\textbf{s}_j$ as its columns.  
The concept of a strategy vector $\textbf{s}_j$ that every species $j \in\{1\ldots k\}$ has in order to harvest resources is central in the model. Each component of $\textbf{s}_j$ represents a strategy for a particular resource, the ensemble of which is characterized by a time-dependent vector $\textbf{r} = \{r_1,r_2,\ldots, r_l\}$ where $l$ is the number of resources. $\textbf{s}_j$ quantifies how much of each resource every individual would like to take out of the resource bath. We consider the concept of a resource component $r_i$ as extremely general: it can refer to a specific molecule, a chemical energy influx, or even to a certain amount of space available in a habitat. Each species has limited energy and time to spend harvesting. Thus, species need to optimize their foraging behavior by choosing how to be efficient as regards different resource consumption~\cite{macarthur1966optimal}. To express the interdependence of resources, one simple way is to fix a norm of the strategy vector to an arbitrary quantity. For convenience we choose $\|\textbf{s}_j\|_2 = 1$, but other norms and values can be chosen. This choice does have a biological significance~\cite{caetano2021evolution}. For now, we will focus on speciation with this fixed choice, but in Sec. \ref{subsec:norm} we will come back to it. 

We assume that the total demand for resources is proportional to $\textbf{s}_j$ and the population size of each species $n_j$. The time dynamics of every resource component~$r_i$ is then described by:
\begin{equation}
\diff{r_i}{t} = -\beta\sum_{j=1\ldots k} s_{ij}n_j + \gamma_i.
\label{eq:resdyn}
\end{equation}

\noindent Here $\beta$~is a timescale, and $\gamma_i$ is the resource replenishment factor of resource~$i$, essentially representing a chemostat~\cite{posfai2017metabolic}. In this simplified linearized resource dynamics, $r_i$ is not strictly positive, which is unphysical. When $r_i<0$ we set it to zero. We find that the time-dependence of the model is very sensitive to the choice of the rate of consumption (and growth). However, the approach of a dynamic equilibrium while adding species to the ecosystem is preserved regardless of the choice of growth rate factor.

If the preferred resource intake of the species is similar to the composition of the resource environment $\textbf{r}$, the growth rate should be maximal; in the case where $\textbf{s}_j$ and $\textbf{r}$ are not aligned, the species should perform poorly. A species goes extinct when its population size is below a threshold value. We verified that threshold choice is not important for much of the dynamics observed. We do note that setting the threshold lower trivially increases the total ecosystem size. It is now natural to write for $n_j$ that
\begin{equation}
\diff{n_j}{t} = \left(\alpha\, \textbf{s}_j\cdot\textbf{r}- \delta\right)n_j.
\label{eq:popdyn}
\end{equation}
Writing explicitly that $\textbf{s}_j\cdot\textbf{r} \equiv \sum_{i=1\ldots l}s_{ij}r_i$ makes clear that $s_{ij}$ is the resource utilization coefficient of species~$j$ for resource~$i$. $\alpha$ is, again, a time constant; $\delta$ sets the population decay rate. Eqns. \ref{eq:resdyn} and \ref{eq:popdyn} are a simplified version of MacArthur equations~\cite{macarthur1970, chesson1990, haygood2002}. However, we interpret the coupling matrix $s_{ij}$ much more specifically: it is essential to see how $\textbf{s}_j$ here serves as the \emph{definition} of species $j$~\cite{posfai2017metabolic}.\\ 

\subsection{Stochastic speciation} 
\begin{figure}[t!]
\centering
\includegraphics[width=0.8\columnwidth]{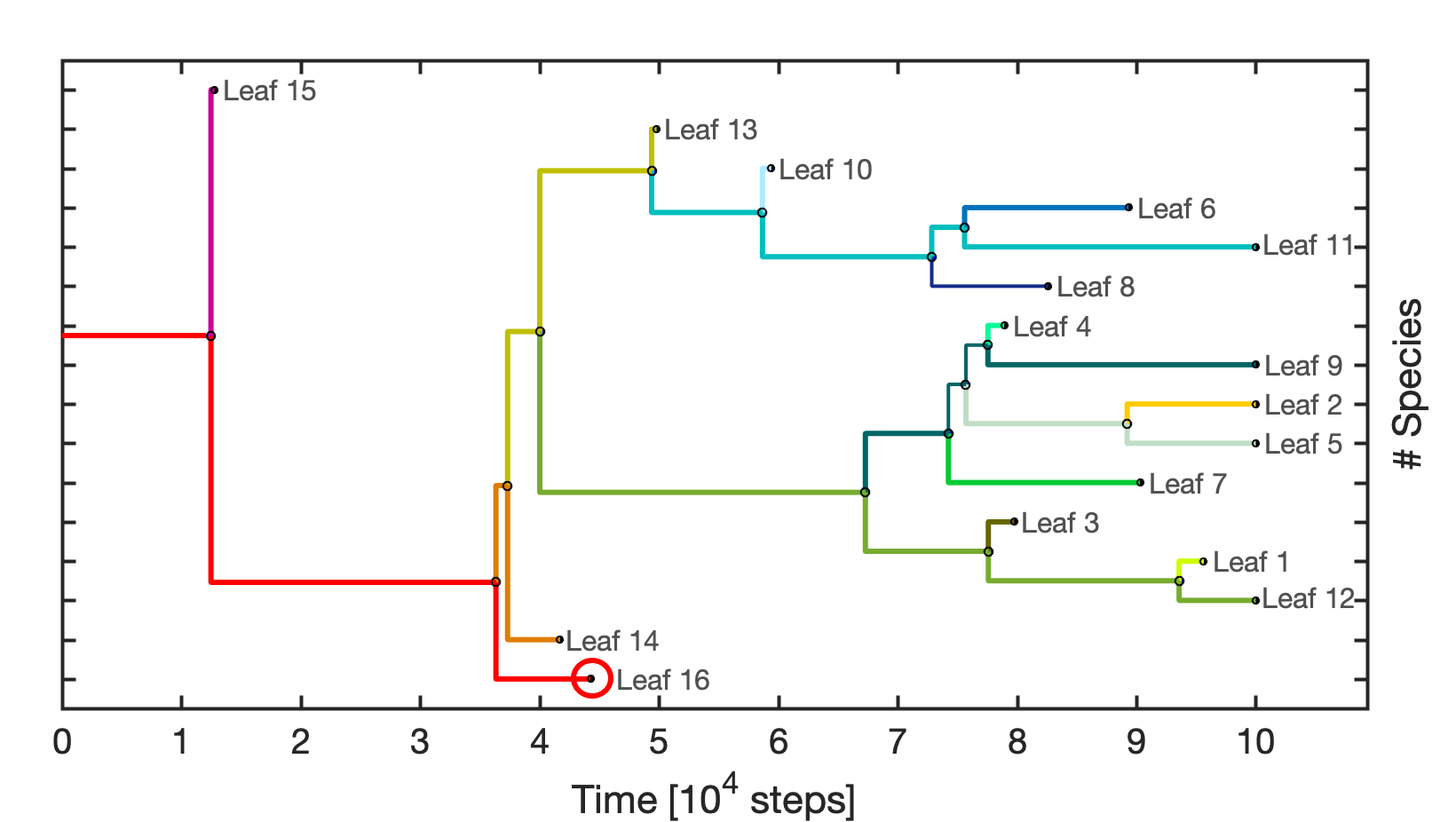}
\caption{The phylogenetic tree derived from the emergence of species in the evolution of one realization. Time is expressed in terms of integration time steps. The species lifetime is indicated in color to provide the chronology of emerged species in the tree. Leaf 1 indicates the emergence of the last spawned species, and so on. The red line shows the lineage of the species with which the tree started; it went extinct around $t \sim 4.4\times10^4$ steps.}\label{fig:1}
\end{figure}

So far, we did not consider any stochasticity: the skeleton of the dynamics is deterministic and embeds a selection for the fittest species~\cite{vellend2010conceptual}. The novel feature in this work is that we add speciation dynamics in two ways: (1) by adding species and (2) by ensuring the added species can be different from existing ones. We choose to make \textit{(i)} mutants appear randomly, adding a new species equation to the system in a Monte Carlo way. We shall see that this captures both evolution and invasion. \textit{(ii)} The strategy vector of the newborn species is stochastically generated ~\cite{posfai2017metabolic, pacciani2020dynamic, drake1990mechanics, servan2018coexistence, may1972will}. Thus, the community assembly happens sequentially at random times with randomly evolved species---see Fig.~\ref{fig:1}. We call our speciating MacArthur approach towards ecosystem dynamics ``SMA'' for brevity.\\

We can now use $\textbf{s}_j$ to specify how a species \textit{evolves} with a simple Monte Carlo evolutionary model: mutants are generated from existing species from which they differ only in terms of the harvesting strategy vector. The initial species in the ecosystem is defined by uniformly drawing an ancestral strategy vector $\textbf{s}_1$, with elements in the range $(0,1)$, and normalizing it to one in the Euclidean norm. A new species can be spawned for every time step and for every alive species, when a random number, drawn from a standard normal distribution, is larger than $\nu$ standard deviations. 
We can thus define a mutant $k+1$ by taking any existing species strategy $\textbf{s}_j$ and by adding a noise vector: 
\begin{equation}
\textbf{s}_{k+1} = \frac{\mid\textbf{s}_j + \eta{\bm \psi}\mid}{\|\textbf{s}_j + \eta{\bm \psi}\|_2}.
\end{equation} 
The noise vector is composed of a random vector drawn from a normal distribution, ${\bm \psi}$, weighted by a parameter $\eta$ quantifying the \emph{amplitude} of the mutation. Therefore, the noise vector represents a shift in the species' resource utilization composition as a consequence of mutations. Note that by so ``phenotypically'' defining our species solely in terms of $\textbf{s}_j$, a natural link to genetic variation within a species is lost. The stochastic arrival of a new species and the resource richness influence the local selection outcome by possibly inducing historical contingency and priority effect in the community assembly~\cite{fukami2015historical, almany2003priority, sale1977maintenance}. The level of historical dependency on species' arrival hinges on the value of $\eta$ and the number of different resource types. We assume that a small $\eta$ represents an infinitesimal evolutionary mutation in a species' survival strategies, originating speciation events in response to the dynamic resource landscape. Indeed, $\eta$ defines the degree of strategies' divergence from parent species to daughter species. On the contrary, foreign invasions are modeled by considering the arrival of a new species with an entirely new set of characteristics, uncorrelated with the ones already present in the system. We call such case $\eta \to \infty$ and will be discussed more in detail in Sec.~\ref{sec:eta}. As we will see, SMA does show a remarkable ability to reproduce behaviors that can be interpreted in any eco-evolutionary context. We will discuss the interpretation of $\eta$ in more detail in Sec.~\ref{sec:roleeta}.\\

\subsection{Biological example: biofilms}
Despite the sober mathematical formulation, SMA conceptually captures some essential features of ecological systems: in several natural ecosystems, ecological successions are intertwined with populations' adaptation to environmental conditions. Moreover, both native species' evolution and alien species' invasion contribute to determining the community's fate, together with environmental responses or sudden changes. One simple yet effective example is that of microbial communities in \emph{biofilm} formation, where evolutionary and ecological timescales are comparable \cite{hansen2007evolution, goyal2022interactions}. In subaerial biofilms, such as those that grow in monumental buildings, the substrate of stones is firstly colonized by pioneer airborne microbes, which then leads to further successive stages with the subsequent invasions of other microorganisms ~\cite{gorbushina2007life}. The different substrate characteristics and environmental conditions define the dynamic resource richness, which in SMA is rendered via the types of resources and their influx vector~\cite{c2020influence,arino2010effects,caneva2004stone}. Moreover, microorganisms not only can feed on others' metabolic discards, but are also able to evolve rapidly via strains' mutations, which in our model is captured by the signs and mutations in the strategy vectors ~\cite{gorbushina2007life}. As a result, several biofilms quickly show resistance to chemical anti-degradation treatments, often leading to unexpected and new community structures \cite{simoes2009species,gorbushina2007life}.

It is evident that biofilms are much more complex ecosystems than the ones described by SMA. In biofilm formation, both evolutionary speciation and ecological invasion act simultaneously, while in the current version of SMA we consider these processes separately for simplicity. SMA can of course include both effects simultaneously, yet we aim to disentangle the dynamics observed in a simple general framework that captures existing natural systems. Many other interpretations besides biofilms are possible and welcome.

\section{Implementation}\label{sec:imp} 
We run all ecosystems starting from one species, with $\alpha = 0.005$, $\beta = 0.01$, $\delta = 0.1$. We assume that all resources have an equal influx rate given by $\gamma_i=1$. We focus on the case of $\nu=3.8$ and we consider systems where we vary $l$.
The value for $l$ can be representative of several different community scales. For example, when considering microbial communities, 100 or more different resource types are an appropriate choice~\cite{fischbach2007one, fischbach2011eating, tikhonov2017collective}. We then varied $\eta$ within previously defined limits and studied the ecosystem's evolution.

\subsection{Solver}
To integrate the species dynamics in Eq.~\ref{eq:popdyn}, we employ a fourth-order Runge-Kutta (RK4) method with stochastic elements that can only generate one new species per existing one at every time step. We checked that the RK4 accuracy used in all our calculations does not affect the results. We verified that our custom implementation provides similar performance to the standard MATLAB \texttt{ode45} solver for the deterministic $l=1$, $k=1$ case, with $\alpha = 0.05$, $\beta = 0.01$, $\gamma_i = 1$, $\delta = 0.2$. In the stochastic setting, we loop in every time step over the extant species and compute for each species the RK4 step and add a new species when a number drawn from a standard normal distribution is, in absolute value, larger than $\nu$. After the loop over the species, we perform an Euler forward step for the resources dynamics Eq.~\ref{eq:resdyn} where we use the updated values for the species. Then we set all negative values for $r_i$ to zero to ensure positivity of the resources. Representative code is provided on Zenodo~\cite{zenodorepo}.

\subsection{Example ecosystem: $l=5$, $k=1$}
The SMA model can show a wide range of different dynamics, depending on the (initial) number of species $k$ and total number of resources $l$ and other parameters. The phenomenology of speciation embedded in the SMA can however already be observed for starting evolution with the most stringent starting condition of \emph{one} species, that is, $k=1$. 
Note that capturing the emergence of an ecosystem with interacting species from a single primordial reproducing entity is one explicit aim of the current modeling approach. Evolving an ecosystem from a single species is, for example, not possible in the classical MacArthur, Lotka-Volterra (LV) or replicator equation contexts.

In this example we choose an $l=5$ resource space and set $r_i(0)=10$ for all $i$ as initial resource amount available. Due to the presence of multiple species, the vector $\textbf{r}(t)$ will be time-dependent and may not always be aligned with a particular species vector $\textbf{s}_j$. We use $\alpha = 0.005$, $\beta = 0.01$, $\gamma_i = 1$, $\delta = 0.1$, $\nu = 3.7$ and $\eta = 1$ and evolve the system for up to $10^5$ time steps of size $h=0.1$. Note that the value of $\nu$ is intrinsically linked to the choice of $h$ because $h$ also sets the frequency at which new species are generated; we come back to this point in Sec.~\ref{subsec:h-nu}.  
We consider a species extinct if its size is smaller than $0.1n_\mathrm{start}$, where $n_\mathrm{start}$ is the initial size of the population; this threshold effectively captures the role of fluctuations in small populations~\cite{reichenbach2006, parker2009, huang2015}. The specific value for the extinction threshold does not affect the essence of the evolutionary dynamics of $\textbf{r}$ and $\textbf{n}$. For repeated independent ecosystem calculations, we generate a new $\textbf{s}_1$ for every iteration. 

\begin{figure}[p!t]
\centering
\includegraphics[width=0.8\columnwidth]{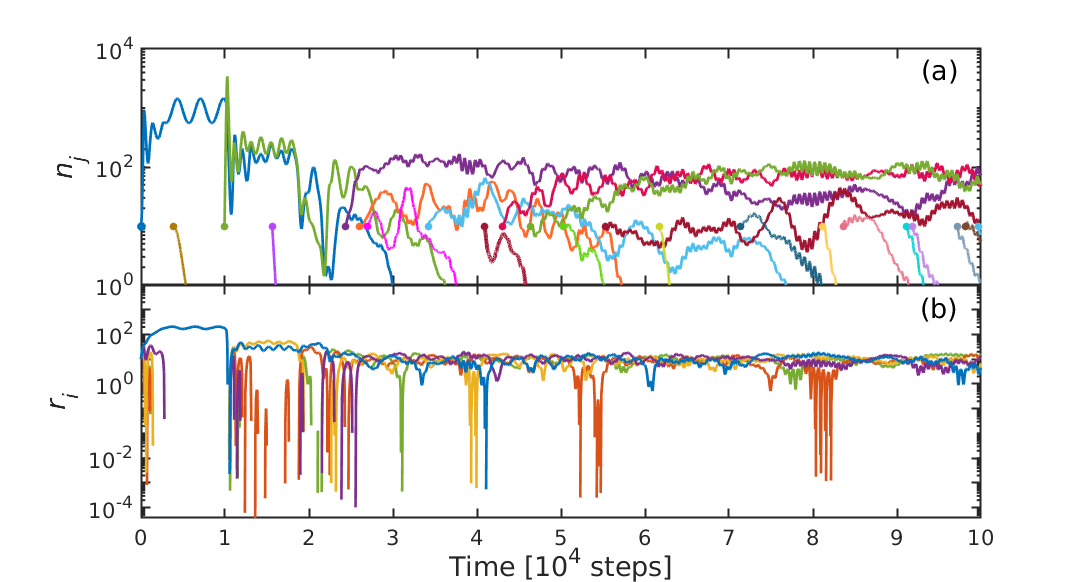}
\caption{(a) $n_j(t)$ for the first (blue) and all subsequently emerged species; color indicates the spawning time, the dot the emergence of a new species. For visual simplicity, we displayed 22 over the 220 species spawned during the community evolution. (b) $r_i(t)$ for the five resources available in the evolving ecosystem. At later times, multiple resources emerge after initial depletion.\label{fig:sm:1}}
\end{figure}

Fig.~\ref{fig:sm:1} shows the typical ecosystem evolution initiated from one species $k=1$ starting at size $n_1 = 10$. Due to the stochastic nature of species emergence and extinction, every realization of ecosystem dynamics is different. However, several important qualitative features reproduce and are visible in this and any example: (\textit{i}) The initial species size oscillates in time until a viable new species has emerged; in sync, the resource dynamics is also oscillatory for the smallest component of $\textbf{s}$ as all the other resources get quenched to zero~\cite{huang2017dynamical}. Note that this excludes the case when an $s_i$ is strictly zero, which is possible but rare. This is such that the resulting behavior dynamically balances the resource usage with the resource influx.  (\textit{ii}) The emergence of new species affects the timescale of periodic oscillations; also new species can make older species go extinct. (\textit{iii}) Later in the evolution, the population fluctuations shift in frequency and decay in amplitude and multiple resources become utilized. The interpretation of these three trends is clear: the randomly selected initial species favors the survival of one resource, for which $s_{i}$ is the smallest. After this transient, the dynamics follow the $l=1$, $k=1$ system which is pseudo-LV in character and allows for periodic orbits of fixed frequency. In this phase, the possibility of the random emergence of new species is consequential: the emergence of new species that are $\eta\bm{\psi}$ different from their parents will suppress the dominant role of the first species and limit its overuse of other resources, thus making the remaining resources emerge again as they are always continuously replenished at rate $\gamma_i$. We will make these statements more quantitative in the next sections.

\section{Main Phenomenology} The first significant result from SMA is the naturally bounded ecosystem it produces in both size and structure while we neither fix the (maximum) number of hosted species in the community nor the maximum population size of the individual species; only the influx of resources $\bm\gamma$ is bounded. In modeling, this is traditionally captured with logistic growth models and/or to restrict oneself to probing the dynamics of an ecosystem with a fixed number of species~\cite{macarthur1970, chesson1990, haygood2002, grover1997resource, posfai2017metabolic, tikhonov2017collective, grilli2017feasibility}. SMA embeds size limitations naturally, as we observe that for enough simulation time, the number of species grows towards a long-term stationary value---see below. This maximum number of coexisting species is solely determined by the distribution of strategies and resource availability. The bounded growth feature allows us to explore the long-term species abundance distribution (SAD)~\cite{hubbell2001unified, mcgill2007species}, with the knowledge that a system will maintain, on average, a constant number of competitors and, as we will see, a finite global size. Note that during equilibrium size, the model allows for and will randomly let species emerge and go extinct; the ceiling represents a dynamic equilibrium. Note that in much of the dynamics explored, $\nu$ mostly sets the speciation rate for the system evolution and is thus a timescale.\\

\begin{figure}[!t]
\centering
\includegraphics[width=0.8\columnwidth]{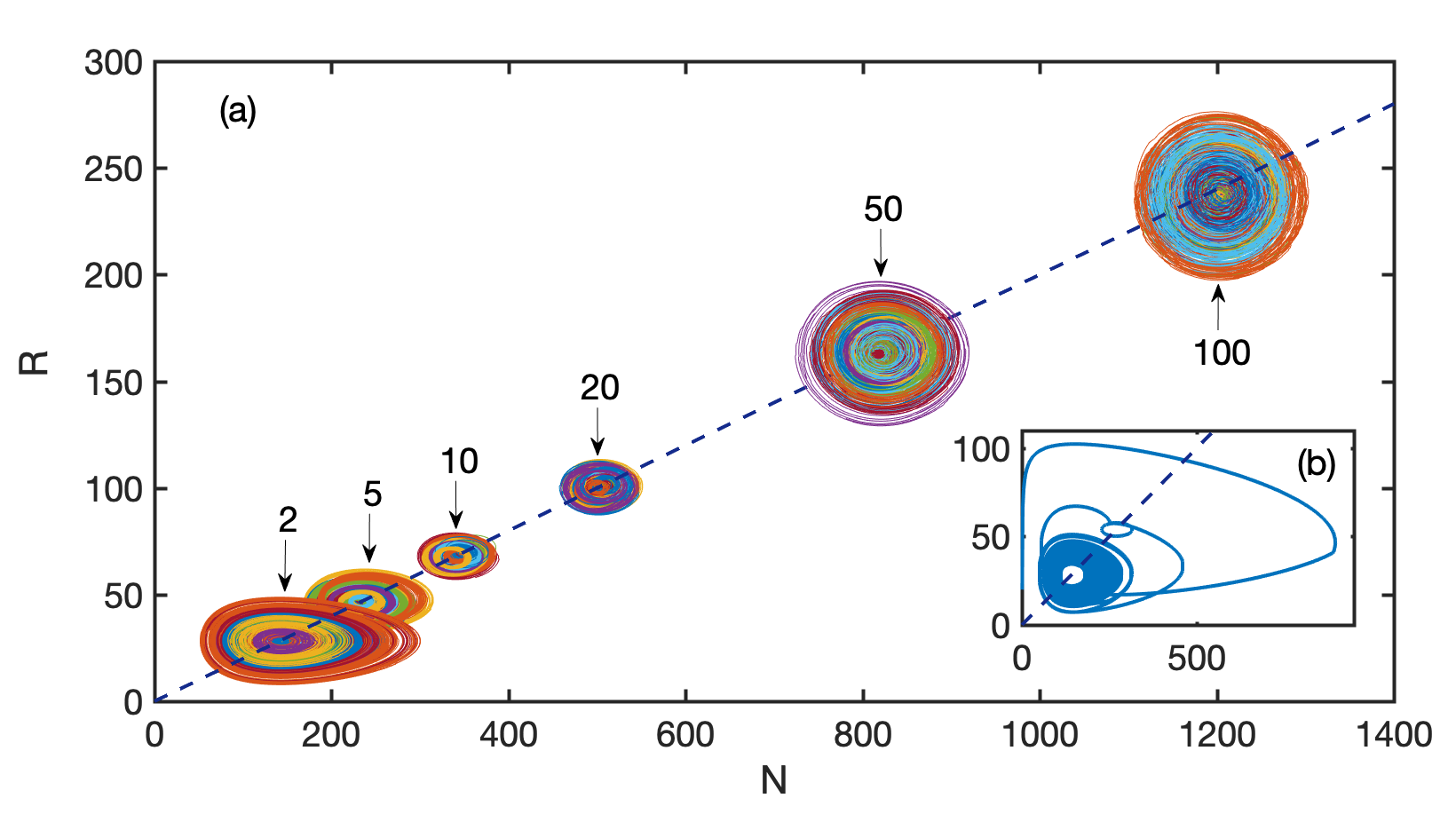}
\caption{ $(R,N)$ phase space of the last $10^4$ out of $10^5$ time steps for 150 realizations (indicated by lines of different colors) with $\nu = 3.8$, $\eta \to \infty$ and different values of $l$ (arrows). For $l=100$ the simulations ended after $8\times 10^4$ time steps due to memory load issues, and the figure displays the last $10^4$ time steps of the runs. The fixed points lie on a line with slope $\frac{l}{\sum \gamma_i}\frac{\delta \beta}{\alpha}$ (black dashed line). (b) Evolution of a single realization in $(R,N)$ space with $l=2$ with initial conditions $R=20$ and $N=10$. The system gradually displays asymptotic limit cycles. Different colors indicate different ecosystem realizations.
}\label{fig:line}
\end{figure}

\subsection{Growth towards equilibrium dynamics}
Rather than considering each species or resource type separately, we gain insight into the system evolution as a whole by considering the total number of individuals ($N$), and the total amount of resources ($R$). Interest in the total abundances dynamics for similar trophic species is seldom suggested and, to our knowledge, rarely explored~\cite{posfai2017metabolic}. Yet, empirical evidence shows that the aggregate biomass could provide valid information on the stability and composition of a community~\cite{tilman1997plant,doak1998statistical}. Thus, studying the evolution of total abundances allows us to explore the system behavior in greater depth from a new perspective and, simultaneously, reduces the variables involved. Resultant equilibrium dynamics for ecosystem averages for different $l$ are shown in Fig.~\ref{fig:line}a. The inset shows how the equilibrium is reached for a particular example setting of $l=2$. We find a family of fixed points for the $(R, N)$ dynamics that are all on a line defined by  

\begin{equation}
\frac{R}{N}=\frac{l}{\sum_i \gamma_i}\frac{\delta \beta}{\alpha}.
\label{eq:line}
\end{equation}

This total abundance dynamics can be understood by simply considering separately the sum of the resources $R$ and the sum of the species $N$. In this way, the dynamically evolving  dimension of the system reduces to a two-dimensional problem

\begin{align}
\label{eq:totRN}
\begin{split}
    \frac{d R}{d t} &= \sum_{i}^{l} \gamma_i - \beta \sum_{i}^{l} \sum_{j}^{k(t)} s_{ij} n_j,\\
    \frac{d N}{d t} &= \alpha \sum_{i}^{l} \sum_{j}^{k(t)} r_i s_{ij} n_j  - \delta \sum_{j}^{k(t)} n_j.
\end{split}
\end{align}

Here, we emphasize that the number of living species, $k$, is a function of time: the equilibrium is dynamic in nature.
The stationary solution $(R^*,N^*)$ of this system should solve the equations
\begin{align}
\begin{split}
    \sum_{i}^{l} \sum_{j}^{k(t)} s_{ij} n_j&= \frac{\sum_{i}^{l} \gamma_i}{ \beta},\\
   \sum_{i}^{l} \sum_{j}^{k(t)} r_i s_{ij} n_j   &= \frac{\delta}{\alpha}N^*,
\end{split}
\end{align}
but note again that the elements that make up $R^*$ and $N^*$ do not have to be stationary.
Inspired by Fig.~\ref{fig:sm:1}, we now assume $r_i$ is approximately constant in $i$, meaning that the mean abundance per resource does not vary too much per resource, we can pull $r_i=\frac{R^*}{l}$ out of the sum, resulting indeed in the fraction

%we see that
%\begin{equation}
%   \left( \frac{\sum_{i}^{l} \sum_{j}^{k(t)} r_i s_{ij} n_j}{\sum_{i}^{l} %\sum_{j}^{k(t)} s_{ij} n_j}   \right) \approx \frac{R}{l}.
%\label{eq:RN}
%\end{equation}
%The ratio between Eq.~\ref{eq:statR} and \ref{eq:statN}, simply becomes
\begin{equation}
    \frac{R^*}{N^*}   = \frac{\delta l \beta }{\alpha \sum_{i} \gamma_i}.
\label{eq:ratio}
\end{equation}

It turns out that this equation predicts the slope of the line in the $(R,N)$ phase space on which all the attractors of the total abundances dynamics lie, as is shown in Fig.~\ref{fig:line}a. Note however that the pictures shown are for a constant $\bm\gamma$, and numerical results seem to indicate that the assumption $r_i\approx\frac{R^*}{l}$ becomes less valid when $\gamma_i\neq\gamma_j$. 
Using the results from \cite{posfai2017metabolic}, we can understand why this line has such predictive power. In the deterministic version of our model, i.e. without speciation, any number of species can coexist. That is, as long as the geometric conditions introduced in \cite{posfai2017metabolic} on the strategy vectors $\mathbf{s}_j$ and the replenishment $\bm \gamma$ are met. When these conditions are met, the system converges to a fixed point where all $r_i$ attain the same value. Hence, what we observe is that every time a new species is introduced (or an old one removed), the dynamics converges to a new fixed point that is indistinguishable from the old one in the $R$-$N$ dynamics. We conclude that the stochasticity in our system always results in an ecosystem where the necessary geometric conditions for the coexistence of many species are met. To be precise, in \cite{posfai2017metabolic} results were obtained for a nonlinear version of our model, such as Eq.~ \ref{eq:newrn}, with normalization in $L^1$ instead of the Euclidean norm. This norm changes some of the details, see Sec.~\ref{subsec:norm}. Also, the nonlinear version of the model changes the slope of Eq.~\ref{eq:ratio} somewhat, while making it valid under more general types of $\bm \gamma$.

 \begin{figure}[p!t]
 \centering
\includegraphics[width=0.8\columnwidth]{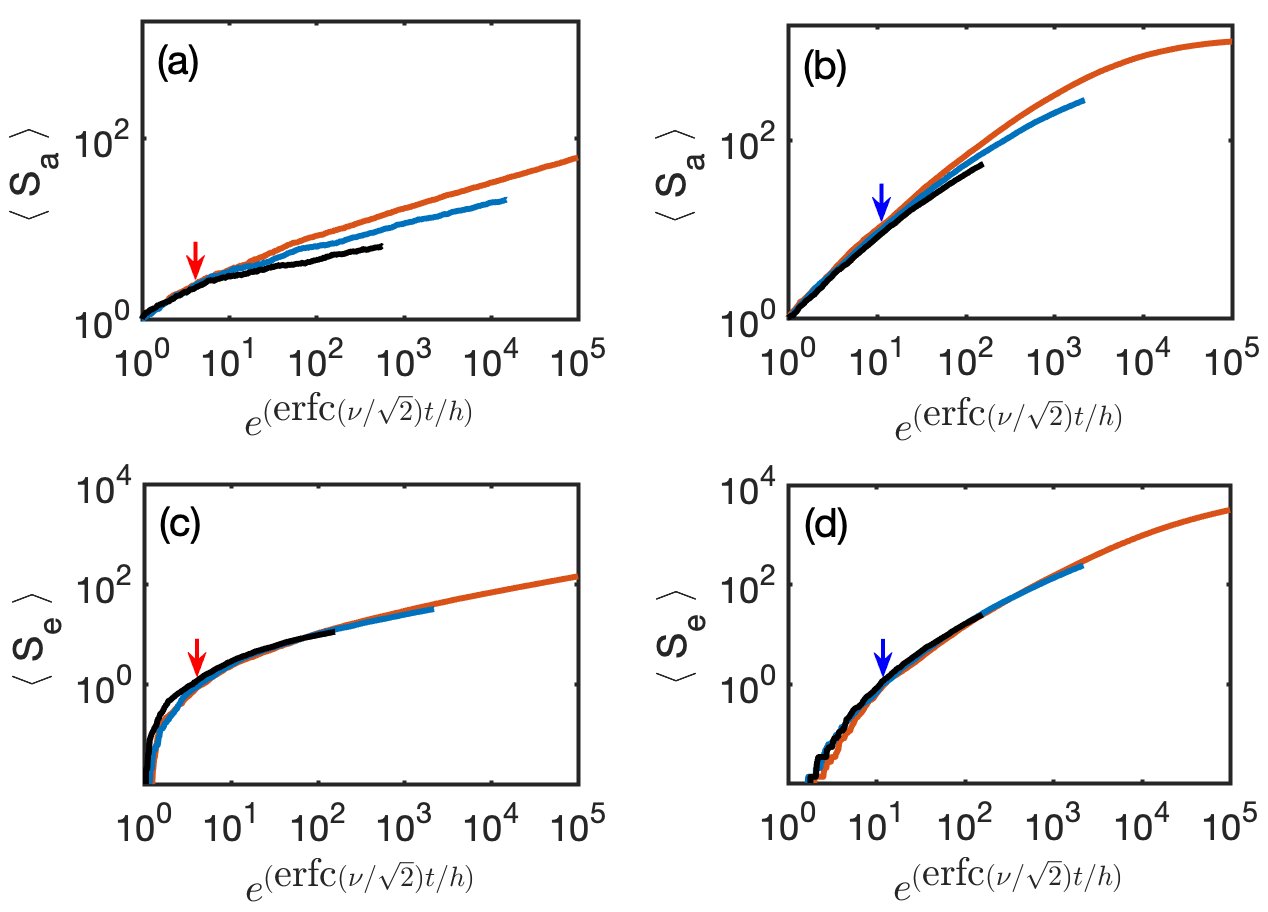}
\caption{(a) Time-scaling of the average number of living species, $\langle S_a \rangle$, over 150 realizations of systems with $l=2$ and $\eta \to \infty$. Curves associated to $\nu=3.8$, $\nu=3.9$ and $\nu=4$ are displayed in orange, blue and black, respectively. The curves are rescaled according to $\tau=\exp(\text{erfc}(\nu/\sqrt{2})t/h)$. The red arrow indicates $\tau \approx 4$, for which the three curves diverge. (b) Same as (a) for a system with $l=100$. The blue arrow indicates $\tau \approx 12$ for which the three curves diverge. (c) Time-scaling of the average extinction, $\langle S_e \rangle$, for the same system as displayed in (a). The colors are consistent with (a). The red arrow indicates the value of $\tau$ for which $\langle S_e \rangle = 1$. For all three different curves $\langle S_e \rangle = 1$ for $\tau \approx 4$, value for which the curves diverge in (a). (d) Same as (c) for the system described in (b). The blue arrow, indicating the value of $\tau$ for which $\langle S_e \rangle = 1$, is consistent with (b) for all three curves ($\tau \approx 12$). \label{fig:nu}}
\end{figure} 

\subsection{Transient scaling with $\nu$}\label{subsec:h-nu} 
A second feature in SMA is that the speciation threshold, $\nu$, induces a timescale $\tau$ for the evolution. At every time step, each species has a probability $p_m$ to mutate. This probability is given by the tail ($\geq \nu$) of the standard normal distribution, i.e. $p_m=\text{erfc}(\nu/\sqrt{2})$.  Therefore, in absence of extinction, we expect the average number of species $\langle S_a\rangle$ to grow as $\sim\exp(\text{erfc}(\nu/\sqrt{2})t)$. However, it should be noted that $\nu$ is intrinsically entangled with the choice of the time step $h$: indeed, $\nu$ defines the probability of an alive species generating a new species in the $h$ time unit. In our simulations, we always kept $h$ constant at the value 0.1 which ensures the stability of the solver. Because of this choice, we must take $\langle S_a\rangle \sim\exp(\text{erfc}(\nu/\sqrt{2})t/h)$.

For $\nu$ to set such a timescale, the average behavior of a dynamic observable obtained for different values of $\nu$, when plotted as a function of $\tau=\exp(\text{erfc}(\nu/\sqrt{2})t/h)$, should collapse into the same master curve. We explored this possibility by considering as observable the time-dependent average of living species $\langle S_a\rangle$ over 150 realizations, keeping $\eta$ and $l$ fixed but setting $\nu=3.8,3.9$ and $4$.

We rescaled the time axis of each curve associated with a different value of $\nu$ to $\tau$. For small $l$ and high values of $\eta$, we observe that $\nu$ does induce a timescale for the overall evolution: indeed, the $\langle S_a \rangle$ curves associated with the different $\nu$ tend to collapse towards a single curve---see Fig.~\ref{fig:nu}a,b in which we display results for $l=2$ and $l=100$ respectively and $\eta\rightarrow \infty$. It follows that for low $\nu$ the evolution is faster, while high values of $\nu$ slow the community's formation.

It is also clear from the figures that $\tau$ is a valid timescale for small communities. When the number of species grows, extinction becomes important and the curves for $\langle S_a \rangle$ deviate weakly from the timescale $\tau$. Quantitatively, when the average number of extinctions $\langle S_e \rangle > 1$ we find that the rescaling becomes less accurate. Curiously, the rescaling works very well for all $\langle S_e \rangle$.

\section{Adaptation} 
The resource alignment interpretation of SMA clearly gives it many physically meaningful links to real world ecosystem dynamics. Empirically, a promising constraint is to provide a time-varying resource influx by introducing~$\gamma_i(t)$, sometimes also called a ``pulse'' experiment~\cite{tilman1987,hiltunen2015environmental}. We demonstrate in what follows that SMA shows adaptation under such conditions. Additionally, we explore how the magnitude of changes in subsequent generations as characterized by $\eta$ affects adaptation dynamics.

\subsection{Rank abundance} 
To demonstrate the effects of a ``pulse'', we focus on the results obtained for a system characterized solely by invasion events, and $l = 100$. We chose to make the resource shock occur at $t_c = 8\times10^{4}$ steps, and we doubled the length of the simulation to provide enough time for the system to respond to the perturbation. For $t < t_c$, the influx rates are $\gamma_i =1$ for all $i$; when $t \geq t_c$, the new $\sum_i\gamma_i$ is three times that before the perturbation. We chose to distribute $75 \%$ of the new ${\bm \gamma}$ among only $25 \%$ of the resources. This abrupt change in resource influx induces adaptation dynamics by the ecosystem. Solving SMA with time-dependent resource influx over several realizations at previously defined $\alpha, \beta, \delta, \nu$, we observe that ecosystems are able to recover from such a resource shock: when the perturbation occurs, there is an initial stage, after which $R(t)$ and $N(t)$ gradually restore their limit-cycles (not shown). However, for the linear model used here, the position of the attractor in the $(R,N)$ phase space changes according to the new resource influx rates. The center of the oscillations do not lie on the line with slope $\frac{l}{\sum_i \gamma_i}\frac{\delta \beta}{\alpha}$ anymore. On the contrary, given the unevenness of the new resource influx vector, the correct slope seems now proportional to the average influxes of the resource types that are not fully depleted, which in general are the ones associated with the highest resource influx. %Nevertheless, we do not deduct such a ratio from any analytical derivation. Rather, it is hypothesized by numerical observations.

\begin{figure}[t!]
\centering
\includegraphics[width=0.8\columnwidth]{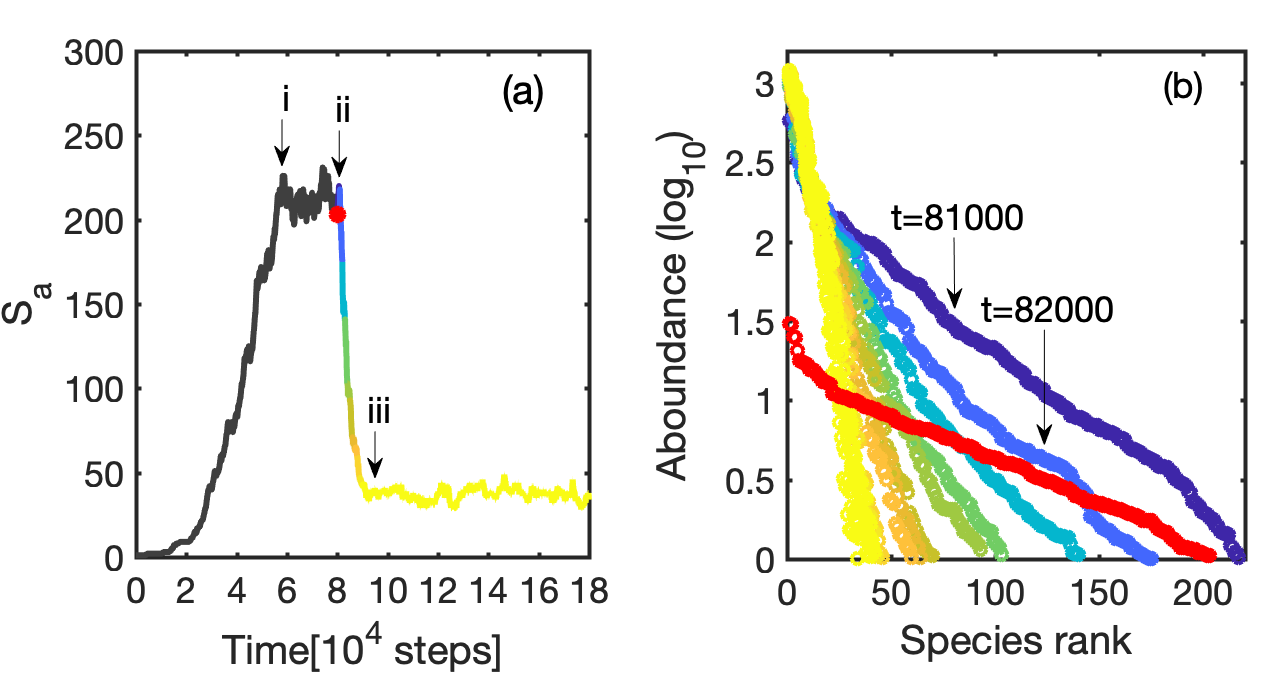}
\caption{(a) The number of living species, $S_a$, for a single realization of a system with $l=100$, $\nu=3.8$ and $\eta \to \infty$. $i$) From $t \approx 4.5\times10^{4}$ steps on, the curve reaches a plateau; $ii$) The resource shock perturbs the system at $t_c = 8\times10^{4}$ steps; $iii$) After the shock, a second plateau is reached. (b) Rank-abundance plot for the realization in (a). The curves show the trend at every 1000 time steps from $t = 8 \times10^{4}$ to $t = 9.1 \times10^{4}$. From $t = 9.1 \times10^{4}$ to the end of the simulation, corresponding to when the $S_a$ reach the second plateau in (a), the Rank-abundance curves are displayed every 14000 time steps. The color scheme follows the colors on the left. The solid red line is the curve when the shock occurs. The arrows indicate the time immediately after the perturbation. 
}\label{fig:4}
\end{figure}

We quantify the pulse response by probing species occurrence. Interestingly, the number of coexisting species is strongly affected by the resource shock. In the absence of perturbations, the number of living species hosted in the system spontaneously grows until it reaches an average maximum value in time---see Fig.~\ref{fig:4}a. When the perturbation occurs, the increase in available resources initially encourages the system to welcome new species, resulting in a sharp peak in the number of coexisting species. Subsequently, the living species curve decays with a characteristic timescale until it reaches a substantially lower new stationary value---see Fig.~\ref{fig:4}a. The new rank abundance distribution corresponds to having fewer species that are all large in population size.

Remarkably, the shock also influences the SAD---see Fig.~\ref{fig:4}b. Before the resource shock occurs, the rank-abundance plot, also known as Whittaker plot~\cite{magurran2013measuring}, displays a curve that gradually collapses towards lognormal-like behavior in the tail. Such behavior resembles that observed in empirical data~\cite{sugihara1980minimal, longino2002ant, baldridge2016extensive,magurran2013measuring, may1975patterns}, although a few methodological aspects that give rise to such distribution are still debated~\cite{magurran2013measuring, may1975patterns}. 
After the perturbation, the curve still preserves its characteristic shape. However, its slope gets steeper in time, and the curve reaches a new asymptotic behavior characterized by less species evenness. Curiously, for systems with $l=2$, the rank-abundance trend shows a strongly uneven species distribution that is often associated with harsh environments or early stages of successions~\cite{magurran2013measuring, mcgill2007species} (not shown). Further generalizations of SMA are discussed in Sec.~\ref{sec:gen}.\\

\begin{figure}[t!]
\centering
\includegraphics[width=0.8\columnwidth]{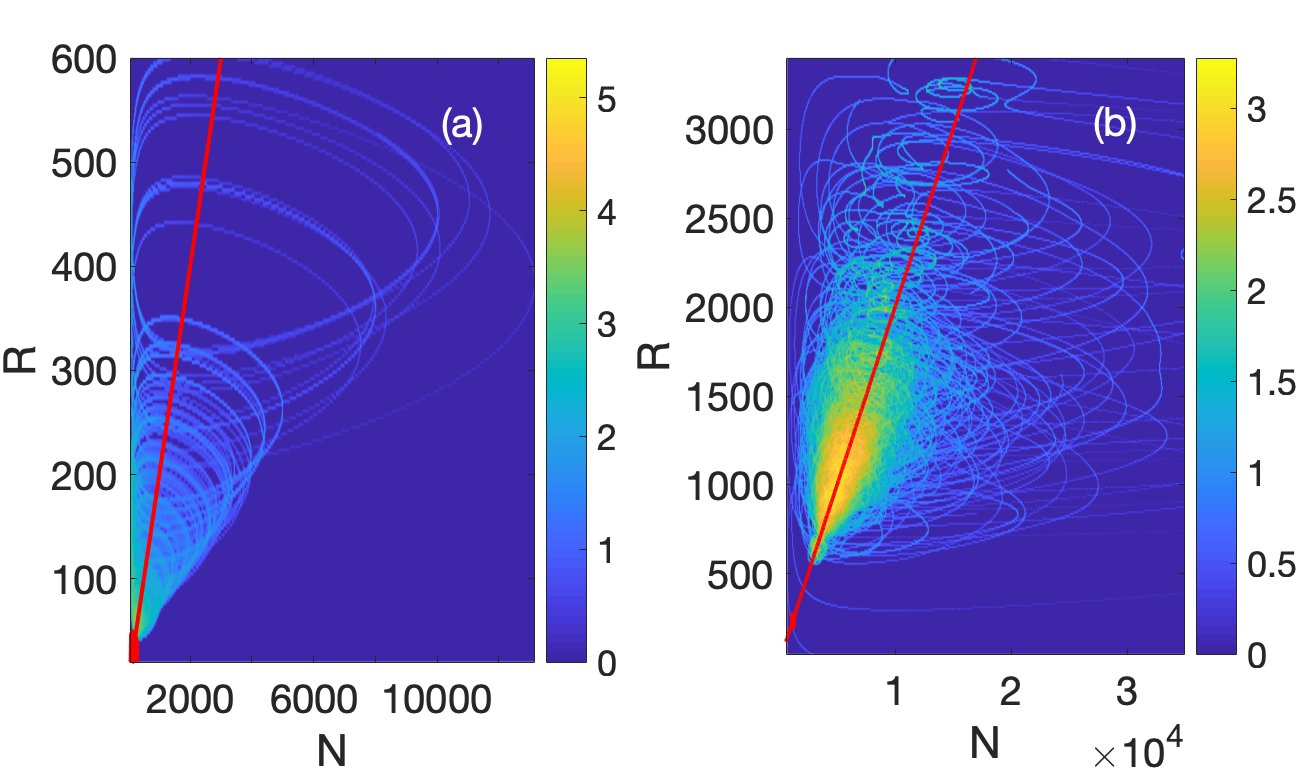}
\caption{(a) $(R,N)$ heatmap of the last $2\times10^4$ out of $10^5$ time steps for 150 realizations with $\nu = 3.8$, $\eta \to 0$, specifically $\eta=0.005$, and $l=2$, colored according to the logarithm of the number of counts. Each realization displays limit cycles around a different fixed point. In red, the line with slope $\frac{l}{\sum \gamma_i}\frac{\delta \beta}{\alpha}$. The red dot on the red line but close to the origin identifies the late time dynamics for $l=2$ and $\eta \to \infty$ as shown in Fig.~\ref{fig:line} in the main text. (b) Same as (a) but with $l=100$. Most of the aperiodic trajectories are in the proximity of the line.}\label{fig:3}
\end{figure}

\subsection{The role of $\eta$}\label{sec:eta}

The noise amplitude $\eta$ has two biologically different limiting cases. For $\eta \to 0$, the community's evolution proceeds via infinitesimal steps, with all the new species occupying the same niche. On the contrary, we can imagine a scenario in which a foreign species invades the community from an external pool. In this case, we assume that the foreign species evolved from a different ancestral species. Thus, we define its strategy by drawing a new ancestral one and adding a noise vector with the maximum noise amplitude $\eta=1$. This approach preserves the biological interpretation of the ancestral strategy and the mutations, originating from two different distributions: uniform and standard normal. For simplicity, we will refer to this scenario with the term $\eta \to \infty$, although mathematically we do not explore the limit of $\eta \rightarrow \infty$.

\subsection{The case of $\eta \to 0$} As discussed, by decreasing $\eta$, the realizations start to depend on their initial conditions. When $\eta$ is infinitesimal, ecosystem dynamics are mainly determined by the ancestor features. However, such development of a neutral community at a species level~\cite{macarthur2016theory} is made dynamic by a priority effect. Simply put, a small $\eta$ is likely to lead to a successful species only if its ancestor was also successful. For small values of $l$, each realization still defines limit cycles around a fixed point. By increasing $l$, the dynamics becomes aperiodic.
When $l$ is small, as expected the family of fixed points lie along the line defined by the ratio $\frac{l}{\sum \gamma_i}\frac{\delta \beta}{\alpha}$. Also for $l = 100$, the dynamics, even though aperiodic, is still contained in a region of the phase space close to the line---see Fig.~\ref{fig:3}.

We conclude that tuning $\eta$ allows us to apply SMA to both evolutionary and invasion-type dynamics. The so embedded co-occurrence of both selection and priority effect mirrors empirical evidence and theoretical hypotheses suggesting that stochasticity and determinism in community assembly work hand in hand~\cite{chase2011disentangling, dumbrell2010relative, luan2020coupling, cavender2009merging, zhou2017stochastic, losos1998contingency}. 
 \\

\subsection{The case of $\eta \to \infty$} For systems invaded by foreign species, the dynamics of several different realizations of one system exhibit the same attractor and qualitative behavior---see Fig.~\ref{fig:line}. From said figure it is clear that at long time scales, the dynamics settles on quasiperiodic orbits around points on the line of fixed points defined by Eq.~\ref{eq:line}, and higher values of $l$ result in higher values of $R$ and $N$. Varying the parameters $\alpha$, $\beta$, $\bm\gamma$, and $\delta$ gives similar results, only changing the slope of the line. Moreover, for high $\eta$, in the range $[0.5, 1]$, the results are similar to those obtained for $\eta \to \infty$. In this limit of $\eta$, especially the large $l$ limit is interesting, because at small $l$, the ecosystem quickly selects the best adjusted strategies, all the others going extinct. For large $l$, species' strategies are constantly evolving towards an existing optimum that is however statistically unlikely to achieve, leading to slow dynamics. The effect of introducing new species is now also determined by their time of arrival, which now defines their competitiveness, inducing a priority effect~\cite{fukami2015historical, almany2003priority, sale1977maintenance} that we will see is the dominant driver of dynamics in the case $\eta \to \infty$. 

\begin{figure}[!t]
\centering
\includegraphics[width=0.8\columnwidth]{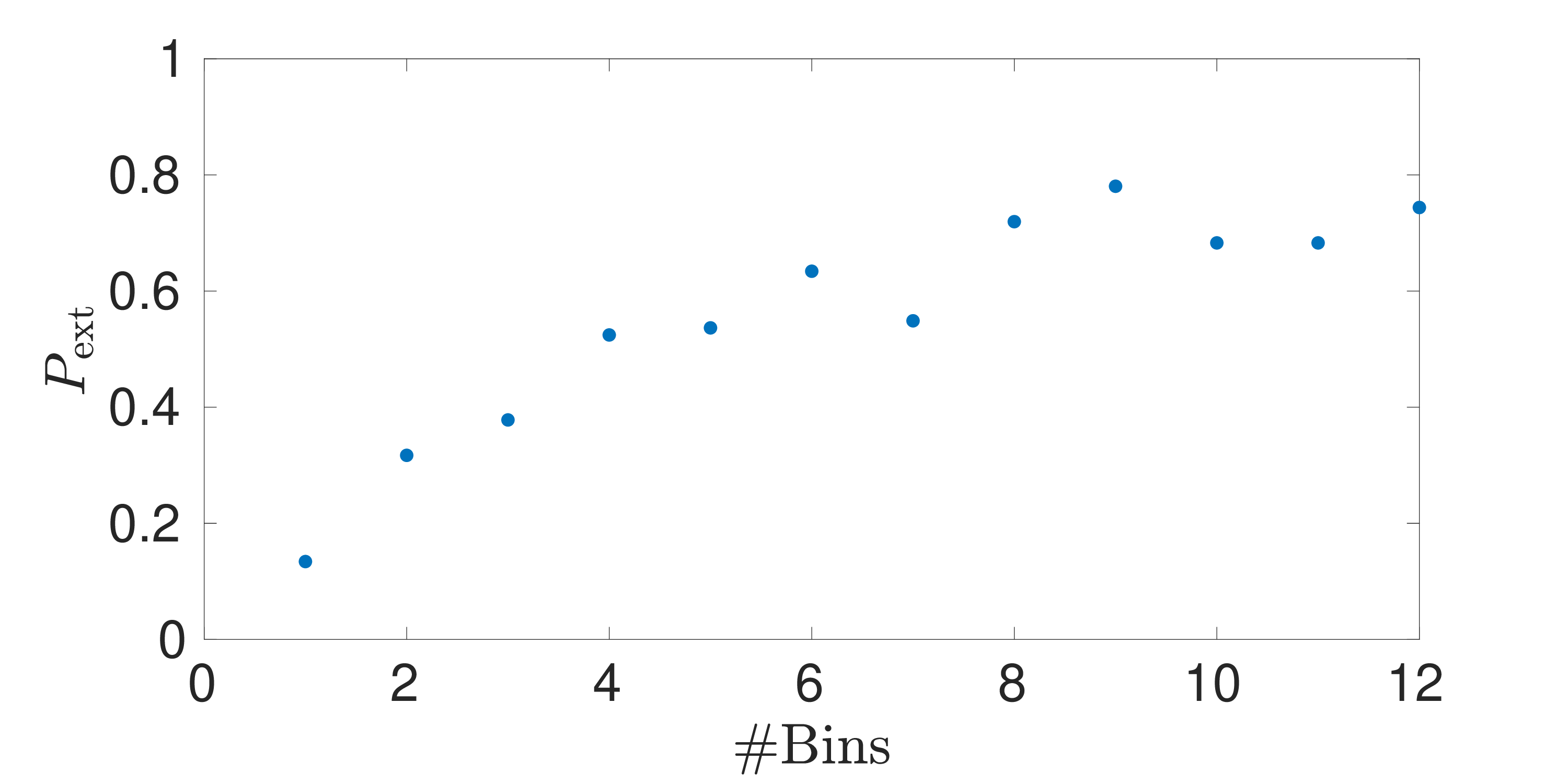}
\caption{Extinction probability for a single realization with $\nu=3.8$, $\eta\to\infty$ and $l=50$. After $6\times10^4$ time steps, 984 species spawned which we divided in 12 bins of size 82. }
\label{fig:sm:Pext}
\end{figure}

\subsection{Priority effect for $\eta\to\infty$}\label{subsec:prior}
In the simulations we see that species that spawn at the end of the simulation have a lower chance of surviving than at the start of the simulation. This can be interpreted as a priority effect. In order to quantify this, we count in a given time interval the number of species that went extinct immediately, i.e. decay exponentially from the initial population to the extinction threshold. Dividing this number by the total number of spawned species in the same interval gives us an immediate extinction probability $P_\mathrm{ext}.$ When the number of spawned species is large enough, we can divide the list of species up into bins and calculate $P_\mathrm{ext}$ for each bin separately which indicates how $P_\mathrm{ext}$ changes over time. Fig.~\ref{fig:sm:Pext} shows a clear trend for an example ecosystem which indicates that there indeed is a priority effect.

\section{Interpreting the role of $\eta$}\label{sec:roleeta}

\subsection{Noise, priority effects and historical contingency}
We observed that the SMA dynamics solely depends on the number of different resources if the community assembly emerges from adding essentially random species. This scenario comes about in the limit of large $\eta$, thus with more significant differentiation between parent and child species, resulting eventually in always the same type of community structure and species' distribution.The late-time community that emerges in this limit is commonly referred to as the \emph{climax community} and indicates the final ecological succession of the community formation ~\cite{morin2009community, weiher1995assembly}. The species, specifically the strategies selected to survive in the climax community, are considered resistant to the invasion of new strategies' variants, which might be considered a type of ``priority effect''.\\ 
On the other limit, we observe that the community assembly is affected by the history of the species' arrival for low values of $\eta$. The community is thus historically contingent ~\cite{morin2009community, belyea1999assembly, schroder2005direct}, an observation that is coherent with the literature. The smaller values of $\eta$ result in more minor divergences between parent and child species. The community is locally neutral because the species belong to similar trophic levels and have comparable survival probabilities. Thus, the community assembly is likely susceptible to the species' arrival history and the pioneer species' features~\cite{fukami2007immigration}. 
%These terminologies commonly describe ecological successions but are recently generalized to eco-evolutionary contexts~\cite{gillespie2004community, fukami2007immigration, losos1998contingency}. Indeed, both invasions and evolutionary histories can lead to historical contingent communities; at the same time, different eco-evolutionary histories can still result in the same climax community~\cite{faillace2022historical,zee2018priority}. In SMA, we allow $\eta$ to vary in a continuum, thus influencing the strength of community assembly dependency on the species' arrival history. The number of resource types also influences the dependency of the community's structure on the assembly history, independently of $\eta$. A highly rich environment with several different resource types favors one species over others similarly competitive ones depending on their arrival time during the community's evolution, inducing priority effects in the assembly---see Sec.~\ref{subsec:prior}.

\subsection{Resource-noise interactions}
What is the influence of evolutionary noise when there is not much room to be different in resource space? What is the role of noise when there are many resources? Clearly, $l$ and $\eta$ span a phase space of dynamics that we briefly explore for the biologically relevant dynamics it can describe.

\begin{figure}[p!t]
\centering
\includegraphics[width=0.8\columnwidth]{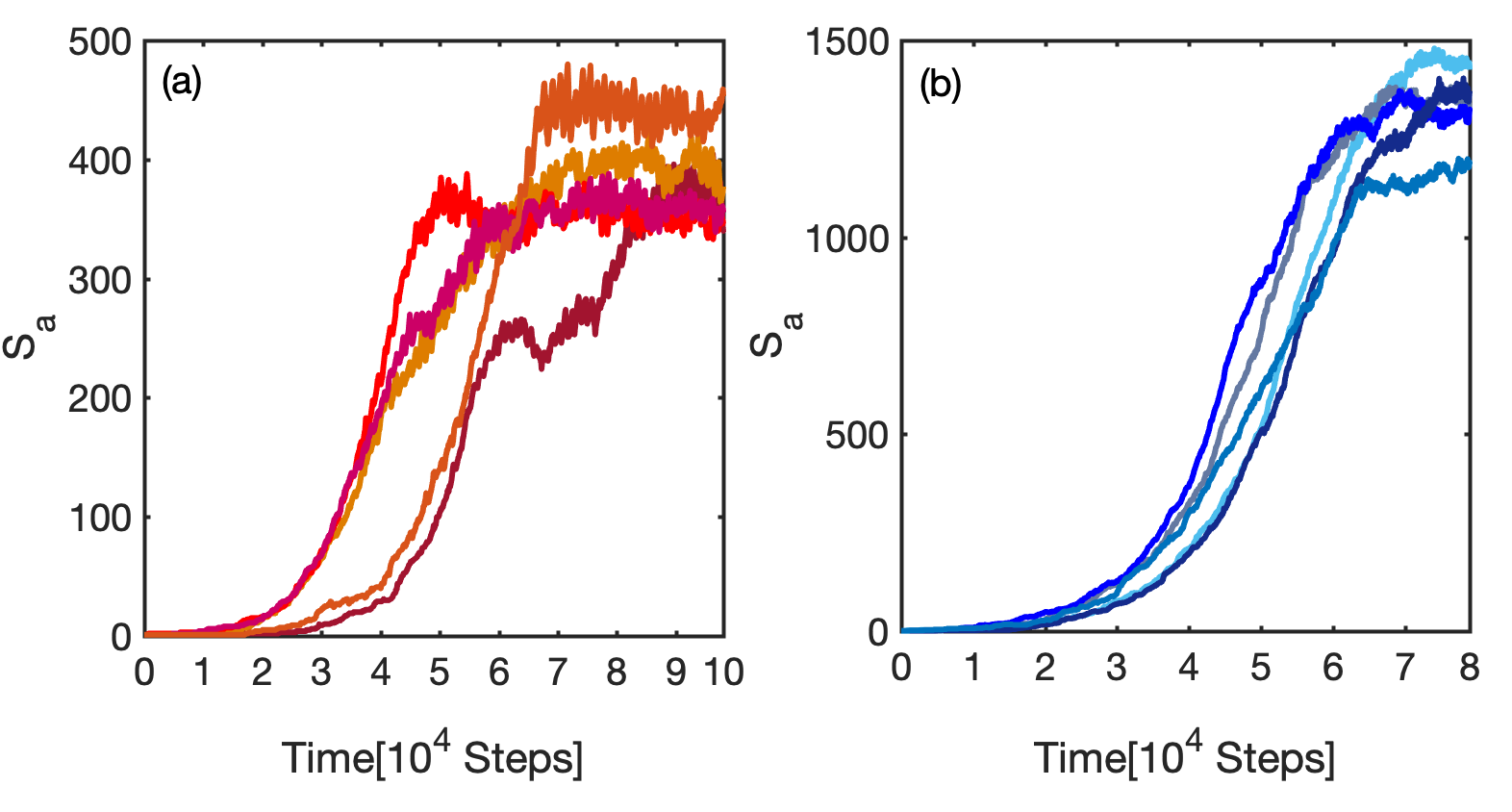}
\caption{(a) The number of living species, $S_a$, for 5 realizations uniformly drawn from a 150 pool of a system with $l=2$, $\nu=3.8$ and $\eta \to 0$. (b) Same as (a) for a system with  $l=100$, $\nu=3.8$ and $\eta \to \infty$. The simulation length of the system in (b) is $8 \times 10^4$ steps due to memory load issues. For both the system in (a) and (b) the $S_a$ curves reaches a long time plateau whose value are different for each realization. \label{fig:plateau}}
\end{figure}

For most realizations of systems with $\eta \to 0$ and $l=2$, the number of living species $S_a$ per realization displays a sigmoidal behavior in time---see Fig.~\ref{fig:plateau}a. This results from the resource-consumer feedback: the total consumers increase by consuming resources before reducing again due to excessive competition and resource lack. However, in this scenario, the species are almost equivalent in their competitiveness, and they can only go extinct if they arrive in the community at an unfavorable time. It seems reasonable to assume that when $l=2$, only a limited pool of strategies can survive in what we can call a ``harsh'' environment. Consequently, the majority of invading species do not present the necessary conditions and quickly go extinct. Successful invasions become rare events, and growth of $S_a$ stalls, reaching an equilibrium. 

Similarly, also systems with $\eta \to \infty$ and $l=100$ show a long-term stationary value for $S_a$---see Fig.~\ref{fig:plateau}b. Here the strong species selection provides additional resource-consumer feedback by choosing the most convenient strategies, leading the less fit species to extinction. Regardless of the interpretation, it is notable that again we observe that long term stability is achieved under very different settings. Note also the much larger number of species that manages to coexist in the limit $\eta \to \infty$ and $l=100$ than when $l$ is small.\\
On the contrary, for the case with $\eta \to 0$, $l=100$ the system does not present any long-term stationary value for the number of living species within the available computation time (data not shown). In this limit, the species take a long time to adjust their size according to the availability of the many resources types.\\ 
In the limit of $\eta \to \infty$ and $l=2$, many realizations of the system hint at the existence of a long time plateau value of $S_a$. Also here, it is reasonable to assume that when $l=2$, a limited pool of strategies can survive in such a harsh environment, but apparently invasion in harsh environments is notably different in its dynamics than evolutionary speciation.

\section{Generalizations of SMA}\label{sec:gen}
SMA allows for many further generalizations~\cite{macarthur1970, chesson1990, haygood2002}. In Eq.~\ref{eq:resdyn} we have only considered that species \emph{consume} resources; however, they may also \emph{provide} resources---see Sec.~\ref{subsec:neg}. Oxygenic photosynthesis~\cite{knoll2017} is one example; on a different scale also gut microbes provide natural resources for each other~\cite{faust2012, vet2018}. Besides, predator dynamics can be introduced by adding another predator coupling matrix term $\sum_im_{ij}n_i$, which can have both positive and negative elements, when species~$j$ is a predator or prey respectively. Growth rates can be made an explicit function of $r_i$, preserving much of the dynamics presented here but adding more biologically relevant constraints. Several other choices can be modified, such as making the speciation rate or $\eta$ a function of $\mathbf{n}$. Making $r_i < 0 $ for some $i$ can account for stressors. We discuss some of these aspects in this section.

\subsection{Negative strategies}\label{subsec:neg}
\begin{figure}[p!t]
\centering
\includegraphics[width=0.8\columnwidth]{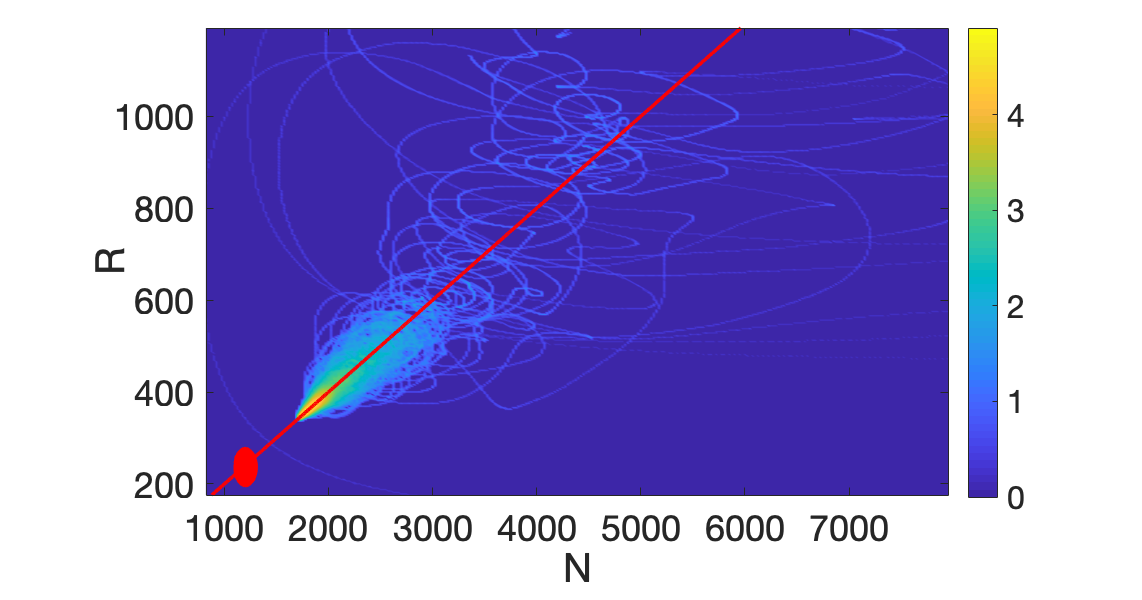}
\caption{ $(R,N)$ heatmap of the last $2 \times 10^4$ time steps for 150 realizations of a system with $l=100$, $\eta \to \infty$ and $\nu=3.8$. Both positive or negative components can define the species strategy vectors, so that $s_{ij} \in \mathbb{R}$. The colormap displays the logarithm of the number of counts. The dynamics are aperiodic but still constrained in a region in the proximity of the line of slope $\frac{l}{\sum \gamma_i}\frac{\delta \beta}{\alpha}$ (solid red line). The red dot indicates the position of the fixed point for a system with only $s_{ij} > 0$.   \label{fig:negstrat}}
\end{figure}

Enabling the components of the strategies $\textbf{s}_j$ to have also negative values is a natural choice or ecological dynamical modeling, as it allows for species to contribute to the resources of other species. 
We find that allowing for the sign change of $\textbf{s}_j$ results in a significantly different transient. For example, when one samples $\textbf{s}_j$ from the full normal distribution, we observe that for infinitesimal mutations, $\eta\rightarrow 0$, and small numbers of resources, $l$, no negative strategy appears or survives. For high $l$ and $\eta$ values, on the contrary, a portion of negative strategies survives. As a result, the total abundance dynamics presents aperiodic behavior, as the negative strategies work as an additional resource influx rate, with an intrinsic stochastic nature given by the random arrival of species with such features---see Fig.~\ref{fig:negstrat}.

\subsection{Consumption and growth rates as Monod functions}\label{sec:monod}
Until now, we assumed that the resource consumption rate $\beta$ was constant, but it can be reasonable to assume that $\beta$ is a function of the resource availability, so $\beta=\beta(r_i)$. The consumption rate then depends on the species opportunity to find and consume resources. It turns out that also for SMA, this consumption rate function is an important factor that determines certain characteristics of the dynamics.\\
To explore the role of the consumption rate function, one relevant choice for $\beta(r_i)$ is the Monod function ~\cite{monod1949}: $\beta_{\rm max} \frac{r_i}{K + r_i}$, where $K$ ($K > 0$) defines the half-saturation constant, that is, the resource availability that is present in the system when the consumption rate reaches half-speed, $\beta=\beta_{\rm max}/2$. The Monod function usually describes bacterial communities' growth dependency on substrate concentration outside the lag phase~\cite{liu2020}. However, we employ it here to express the intuitive concept that the consumption rate will vary depending on the substrate concentration, assuming $\beta(0)=0$, and saturating over a certain level of resource availability $\beta_{\rm max}$. More generally, $K$ could be different for each resource; for simplicity, we set it equal for all the resource types.

\begin{figure}[p!t]
\centering
\includegraphics[width=0.8\columnwidth]{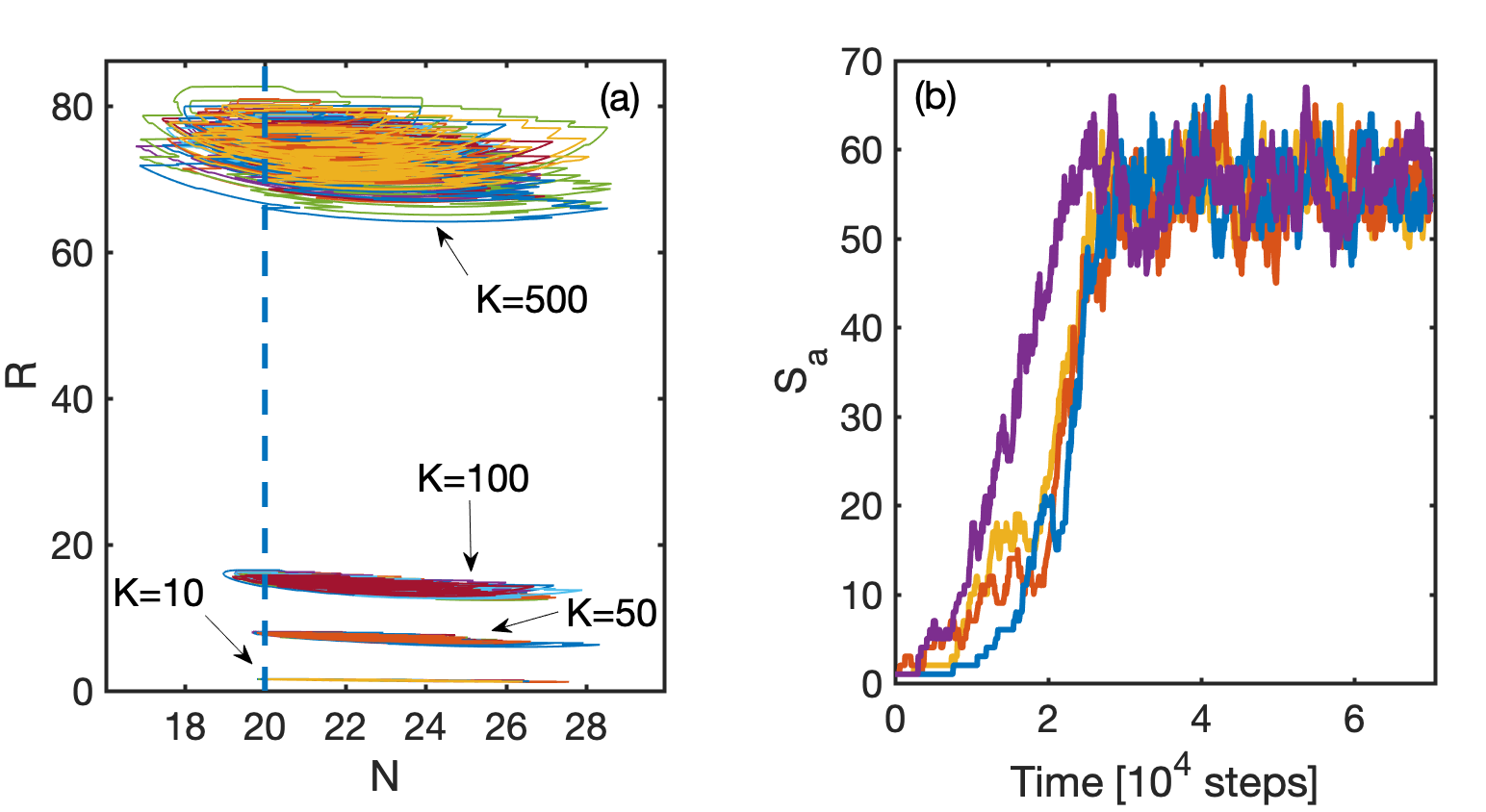}
\caption{(a) $(R,N)$ phase space of the last $2 \times 10^4$ time steps for 20 realizations (different colors) of a system with $\nu = 3.6$, $\eta \to \infty$, $l=2$ and different values of $K$(arrows). Both consumption and species growth rates are described by a Monod function. Here, similarly to ~\cite{posfai2017metabolic}, $\alpha_{\rm max}$ and $\beta_{\rm max}$ were set both to $1$. The attractors lie on a vertical line $ N=\frac{\alpha_{\rm max} \sum_i\gamma_i}{\delta \beta_{\rm max}}$.
(b) Examples of living species curve, one for each system with different $K$, as described in (a).\label{fig:REF2M}}
\end{figure}

Similar to many other resource-consumer models, we further assume that resource uptake by the species is, up to a constant, equal to the depletion of the resources ~\cite{posfai2017metabolic, pacciani2020dynamic}. Consequently, $\alpha(r_i)$ in Eq.~\ref{eq:popdyn} of the main text also becomes a Monod function, differing from $\beta(r_i)$ only on the proportionality constant $\alpha_{\rm max}$: $\alpha(r_i)= \alpha_{\rm max} \frac{r_i}{K + r_i}$.

Summarizing, it follows that the equation for the system dynamics now becomes
\begin{subequations}
\label{eq:newrn}
\begin{align}
\label{eq:newrna}
\frac{d r_i}{d t} &= -\beta(r_i)\sum_{j=1\ldots k} s_{ij}n_j + \gamma_i,\\
\frac{d n_j}{d t} &= \left(  \mathbf{s}_j\cdot \alpha(\mathbf{r}) - \delta\right)n_j.
\label{eq:newrnb}
\end{align}
\end{subequations}

Concretely, we explore the scenario in which $\eta \to \infty$ and varied $l$ and $K$. The choice of $K$ affects both the resource depletion and the species growth times-scales.
The dynamics in $(R,N)$ space does not display limit-cycles as observed for the linearized resource dynamics defined by the systems described in Eq.~\ref{eq:resdyn} and \ref{eq:popdyn} (see Fig.~\ref{fig:sm:1}). However, in the Monod-version of SMA, the attractors still lie along a vertical line, now defined by $N^*= \frac{\alpha_{\rm max} \sum_i\gamma_i}{\delta \beta_{\rm max}}$, obtained by the total abundances dynamics stationary solution. This can be observed in Fig.~\ref{fig:REF2M}a for $l=2$ and Fig.~\ref{fig:monod100}a for $l=100$.  

The long term stationary state reached is also reproduced by the Monod version of SMA. In Fig.~\ref{fig:REF2M}b and  Fig.~\ref{fig:monod100}b for both cases $l=2,100$, we see that independently of $K$, all the systems reach a long time stationary value. In fact, the $S_a$ associated to $l=2$ and $\eta \to \infty$ reach a plateau. This might result from the fact that when a Monod function describes both consumption and growth rates, the systems with low $\nu$ are still computable, while in the linearized version of the model, a low $\nu$ value leads to a too large number of species for the system to remain computationally tractable (but would presumably otherwise reach equilibrium). We see that, a long time plateau is reached for many parameter choices and functional implementations of SMA, strongly suggesting that this feature is a robust property of the SMA model proposed. Of course, the obtained plateau values do depend on the choice of model details.\\ 

\subsection{The role of the resource norm}\label{subsec:norm}
\begin{figure}[p!t]
\centering
\includegraphics[width=0.8\columnwidth]{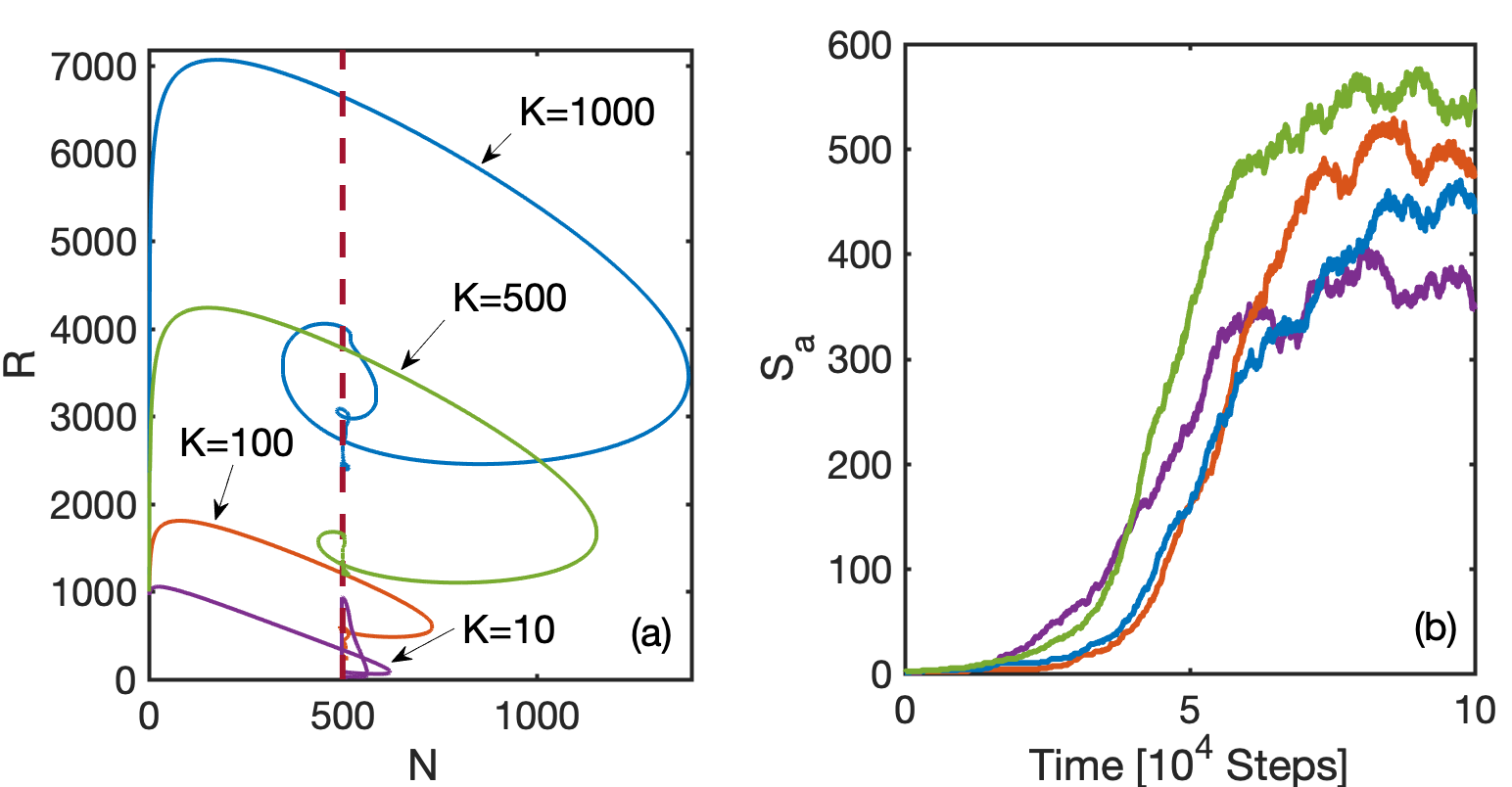}
\caption{$(R,N)$ phase space of the entire evolution of exemplifying realizations (different colors), one for each system with $\nu = 3.8$, $\eta \to \infty$, $l=100$ and different values of $K$(arrows). Both consumption and species growth rates are described by a Monod function. Here, $\alpha_{\rm max}=0.5$ and $\beta_{\rm max}=1$. We notice a fast evolution towards the attractor. (b) Examples of living species curve, one for each system with different $K$, as described in (a). All the systems reach a long time stationary value.\label{fig:monod100}}
\end{figure}

To study the dynamics along the vertical line of attractors, as observed in Fig.~\ref{fig:monod100}a, we have to make a connection with the results in~\cite{caetano2021evolution}. Therefore, let us assume that the strategy vectors are normalized in $L^p$ in stead of just $L^2$. Furthermore, let us consider two extreme options for the structure of $\mathbf{s}_j$: \emph{(i)} all species specialize into consuming one resource and \emph{(ii)} all species have an identical strategy vector $\mathbf{s}^*$ with nonzero components. 

For notational convenience, we write $g(r_i)$ for the Monod function and drop the $max$ subscript in $\alpha$ and $\beta$. This choice will also highlight that the following results do not depend explicitly on the Monod function, but work for any choice of $g$ that is monotone and passes through 0.

From \cite{posfai2017metabolic}, we know the strategy $\mathbf{s}^*$ in $L^1$, so inspired by their proof, we make the following computation. From Eq.~\ref{eq:newrnb} we find that, in equilibrium, we must have
\begin{align}
    \sum_i \mathbf{s}^*_i\alpha g(r^*_i)=\delta. 
\end{align}
We can rewrite this as 
\begin{align}
    \sum_i (\mathbf{s}^*_i)^p\frac{(\mathbf{s}^*_i)^{1-p}\alpha g(r^*_i)}{\delta}=1.
\end{align}
We can read this equation as the inner product of the known vector $(\mathbf{s}^*)^p$ with the unknown vector $\frac{(\mathbf{s}^*)^{1-p}\alpha g(\mathbf{r}^*)}{\delta}$. As we have just one equation to define this $l$-dimensional vector, we have an $l-1$ dimensional solution space. To close the equation, we pick the solution that minimizes the $L^p$ norm. By definition, $\sum_i (\mathbf{s}^*_i)^p=1$, so the most straightforward solution is for every component to be unity:
\begin{align}
\label{eq:minLp}
     \frac{(\mathbf{s}^*_i)^{1-p}\alpha g(r_i)}{\delta}=1.
\end{align} For $p>1$, it is also the unique positive solution that minimizes the $L^p$ norm.   
From Eq.~\ref{eq:newrna}, we deduce that in equilibrium we have the equality
\begin{align}
\label{eq:gsn}
    \beta g(r^*_i)\mathbf{s}^*_iN^*=\gamma_i,
\end{align}
so when we fill in the value of $N^*$ and the result from Eq.~\ref{eq:minLp}, we find that
\begin{align}
    \mathbf{s}^*_i=\left(\frac{\gamma_i}{\sum_i\gamma_i}\right)^{1/p}, \hspace{5mm} p>1.
\end{align}
This is identical to the strategy found in \cite{caetano2021evolution}.

In the simulations, such as Fig.~\ref{fig:monod100}, we noted that the dynamics leads to the minimization of $R^*$, so let us investigate which strategy minimizes $R^*$.
When we invert $g(r^*_i)$ in Eq.~\ref{eq:gsn}, we can find $R^*$ by summing over the individual $r^*_i$ and find
\begin{align}
    R^*=\sum_ig^{-1}\left(\frac{\delta}{\alpha}\left(\frac{\gamma_i}{\sum_i\gamma_i}\right)^{1-1/p}\right),
\end{align}
which is in line with the numerical results. From this equation we learn two things. First, when we choose our normalization in $L^1$, all $r_i^*$ will become independent of $\bm\gamma$, explaining the results from \cite{posfai2017metabolic}. Second, when $p<1$, we see that all $r_i^*$ become smaller when we choose specialization as the preferred strategy. This is again in line with the results from \cite{caetano2021evolution}.

\subsection{Limitations}
After studying several extensions, we must look at the limitations of our approach. We noted that the deterministic version of our model can have an arbitrary large number of coexisting species in absence of an extinction threshold \cite{posfai2017metabolic}. Unfortunately, this is not a generic property of the model. Indeed, when we perturb the constraint that all species have their strategy normalized to the same value, i.e. by letting the value of the normalization depend on the species, we observe that the complex communities collapse back towards a number of species that is in line with the competitive exclusion principle. It is however very well possible that in, for example, plankton communities, many species have similar functional traits \cite{borics2021freshwater,graco2021clumpy}. Therefore, our model might not be the full description of a complex ecosystem, but it can give us insight in the dynamics in a single `cluster' of similar traits.

\section{Conclusions}\label{sec13}
We showed that adding speciation in a MacArthur model adds a host of dynamical features reminiscent of commonly observed evolutionary biology. Even when one starts the dynamics with a single species, by introducing new species of slightly different type into an existing ecosystem, we observe equilibration to a maximum number of species on a time scale that is a simple function of the spawning rate. We observe that the system self-maximizes the number of coexisting species, reaching a long-term stationary value. The stationary behavior represents a dynamic equilibrium as attractor in $(R,N)$ space. Parameters that set the stochastic strength allow the model to explore both invasive and evolutionary dynamics; the size of the resource pool affects the dynamical ability to converge to a niche community~\cite{fisher2014transition}. Community aggregate behavior is also stable under perturbations: the system adjusts its features after a resource influx shock; rank abundance plots are in line with commonly observed features. Much phenomenology is robust when different choices are made for resource consumption rates and other model features and some analytical features of the model are consistent with literature results. The perspective embedded in SMA and the range of biologically relevant phenomena it produces offer a flexible interpretation of the term ``species'' that gives a simple computational tool for a more quantitative understanding of evolution and ecology. One significant open question for this framework is whether the completely random speciation introduced here can also result in the emergence of ``clusters'' of similar species vectors~\cite{maynard2018network}. Such clustering of species that consume resources in a complementary way would be a computationally tractable representation of a true ``tangled bank''. 

\backmatter

%\bmhead{Supplementary information}
%If your article has accompanying supplementary file/s please state so here. 
%Please refer to Journal-level guidance for any specific requirements.

\bmhead{Acknowledgments}
We thank Peter van Heijster, Christian Fleck, Kirsty Wan, Oskar Hallatschek, Arjan de Visser and Peter de Ruiter for various stimulating discussions.

\section*{Declarations}

%Some journals require declarations to be submitted in a standardised format. Please check the Instructions for Authors of the journal to which you are submitting to see if you need to complete this section. If yes, your manuscript must contain the following sections under the heading `Declarations':

\begin{itemize}
\item Funding - The authors thank Wageningen University for supporting this work, and declare that no other funds, grants, or other external support was received for the preparation of this manuscript.
\item Competing interests - 
The authors have no relevant financial or non-financial interests to disclose.
\item Availability of data, code and materials - Representative code used to generate the results will be deposited in a public repository. Data and materials used for the study are available.
\item Author contributions - All authors developed variants of the numerical codes used, performed simulations and interpreted data. E.B. and C.H. derived the main mathematical results, while J.A.D. conceived of the study. All authors contributed to writing the manuscript.
\end{itemize}

\noindent

\pagebreak

\bibliography{merce}
%% if required, the content of .bbl file can be included here once bbl is generated
%%\input sn-article.bbl

%% Default %%
%%\input sn-sample-bib.tex%

\end{document}